\documentclass[12pt]{article}
\pdfoutput=1
\usepackage{subfigure}
\usepackage{amssymb,amsmath}
\usepackage{graphicx}
\usepackage{color}
\usepackage[colorlinks=true
,urlcolor=blue
,citecolor=blue
,linkcolor=blue
,pagecolor=blue
,linktocpage=true
,pdfproducer=medialab
]{hyperref}
\usepackage[a4paper,width=15.2cm]{geometry}
\makeatletter \renewcommand{\@dotsep}{10000} \makeatother

\newcommand{\beq}{\begin{equation}}
\newcommand{\eeq}{\end{equation}}
\newcommand{\bea}{\begin{eqnarray}}
\newcommand{\eea}{\end{eqnarray}}

\begin{document}
%Remove date before submitting to arXiv
%\date{\today}

\begin{center}

 {\Large\bf    A Predictive Yukawa Unified SO(10) Model: Higgs and Sparticle Masses
  } \vspace{1cm}

{\large   M. Adeel Ajaib\footnote{ E-mail: adeel@udel.edu}, Ilia Gogoladze\footnote{E-mail: ilia@bartol.udel.edu\\
\hspace*{0.5cm} On  leave of absence from: Andronikashvili Institute
of Physics,  Tbilisi, Georgia.},  Qaisar Shafi\footnote{ E-mail:
shafi@bartol.udel.edu} and Cem Salih $\ddot{\rm U}$n \footnote{
E-mail: cemsalihun@bartol.udel.edu}} \vspace{.5cm}

{\baselineskip 20pt \it
Bartol Research Institute, Department of Physics and Astronomy, \\
University of Delaware, Newark, DE 19716, USA  } \vspace{.5cm}

\vspace{1.5cm}
 {\bf Abstract}
\end{center}

We revisit a class of supersymmetric SO(10) models with $t$-$b$-$\tau$ Yukawa coupling unification condition,  with emphasis on the prediction of the Higgs mass.  We discuss qualitative features in this model that lead to a Higgs mass prediction close to 125 GeV. We show this with two distinct computing packages, Isajet and SuSpect, and also show that they yield similar global features in the parameter space of this model. We find that $t$-$b$-$\tau$ Yukawa coupling unification prefers values of the CP-odd Higgs mass $m_{A}$ to be around 600 GeV, with all colored sparticle masses above 3 TeV. We also briefly discuss prospects for testing this scenario with the ongoing and planned direct dark matter detection experiments. In this class of models with $t$-$b$-$\tau $  Yukawa unification, the neutralino dark matter particle is heavy $(m_{\tilde{\chi}_1^{0}} \gtrsim 400 \rm \ GeV)$, which coannihilates with a stau to yield the correct relic abundance.

\newpage

%%%%%%%%%%%%%%%%%%%%%%%%%%%%%%%%%%%%%%%%%%%%%%%%%%%%%%%%%%%%
\renewcommand{\thefootnote}{\arabic{footnote}}
\setcounter{footnote}{0}

%%%%%%%%%%%%%%%%%%%%%%%%%%%%%%%%%%%%%%%%%%%%%%%%%%%%%%%%%%%%%

%\baselineskip 36pt
% Main body
%%%%%%%%%%%%%%%%%%%%%%%%%%
%\baselineskip 18pt
%%%%%%%%%%%%%%%%%%%%%%%%%%

\section{\label{ch:introduction}Introduction}

Despite no signals from direct  Supersymmetry (SUSY) searches at the LHC, a 125 GeV Higgs boson is a strong indication that a SUSY signal might be imminent. SUSY still remains at the forefront of beyond the standard model physics scenarios due to several reasons. In addition to solving the gauge hierarchy problem and providing a dark matter candidate, it also leads to unification of the gauge couplings.
In the SUSY case, $t$-$b$-$\tau$ Yukawa unification (YU)  \cite{big-422} can also be accommodated in  contrast to its non-SUSY version.
Both these observations hint {at} an underlying grand unified structure like SO(10) which {may} be supersymmetric.
 The implications of   $t$-$b$-$\tau$ YU has been extensively explored over the years \cite{bigger-422,Gogoladze:2010fu}.

The discovery of the Higgs boson was announced by the ATLAS and the CMS collaborations using the combined 7~TeV and 8~TeV data, with ATLAS observing a  $5.0\sigma$ signal {at} $126.0 \pm0.4 ({\rm stat})\pm 0.4({\rm syst})~{\rm GeV}$ ~\cite{:2012gk},
and CMS a $5.0\sigma$ signal {at} $125.3\pm 0.4 ({\rm stat})\pm 0.5({\rm syst})~{\rm GeV}$ ~\cite{:2012gu}.
This observation was also confirmed at the de Moriond 2013 conference. Spurred by these {exciting} results, we revisit a previous analysis done in ref. \cite{Gogoladze:2011aa} with special emphasis on the prediction of the Higgs mass in {a class of} SO(10) models with  $t$-$b$-$\tau$ YU   condition.
 It was shown {in \cite{Gogoladze:2011aa}} that with non-universal soft supersymmety breaking (SSB)  gaugino masses, which can be derived in the framework of SO(10) GUT,  and with universal SSB scalar Higgs doublet  masses at $M_{{\rm GUT}}$ ($M_{H_{u}}^{2}=M_{H_{d}}^{2}$), {the mass of CP-even SM-like Higgs boson} can be predicted {with} $t$-$b$-$\tau$ YU case. In this paper we {shed light on} the qualitative features that lead to consistency of $t$-$b$-$\tau $ YU with a 125 GeV Higgs.
 Furthermore, we employ two different computing packages, namely Isajet 7.84 and SuSpect 2.41, and show that they agree very well with {the} 125 GeV Higgs mass prediction for $t$-$b$-$\tau$ YU. Good agreement is found between these two programs over the entire parameter space, {thereby} rendering our conclusions more robust.

Another motivation to revisit the analysis presented in ref. \cite{Gogoladze:2011aa} is the recent discovery of $B_s\to\mu^+\mu^-$ {decay} by the LHCb collaboration  \cite{Aaij:2012nna}. {The branching fraction $BF(B_s\to\mu^+\mu^- )=3.2^{+1.5}_{-1.2}\times
10^{-9}$ is in accord with the SM prediction of $(3.2\pm 0.2)\times
10^{-9}$ \cite{Buchalla:1995vs}.}  In SUSY models, this flavor-changing decay {receives contributions from} the exchange of the pseudoscalar Higgs boson $A$ \cite{Choudhury:1998ze}, which is proportional to
$(\tan\beta)^6/ m_A^4$. Since $\tan\beta\thickapprox 47$ was predicted  in ref. \cite{Gogoladze:2011aa}, it is interesting to see how {the} parameter space {is impacted} with the $B_s\to\mu^+\mu^-$ discovery.

{We also} intend to highlight another interesting feature of this YU model, namely that  in addition to the prediction of a CP even Higgs boson mass of around 125 GeV , the model also prefers the CP odd Higgs boson mass of around 600 GeV. This prediction can hopefully be tested at the LHC in the near future \cite{Chatrchyan:2012vp}. {The} colored sparticle masses, consistent with good ($10\%$ or better) $t$-$b$-$\tau$ YU, lie well above the current mass limits from the LHC, i.e., $m_{\tilde{g}} \gtrsim  1.4$~TeV (for $m_{\tilde{g}}\sim m_{\tilde{q}}$) and
$m_{\tilde{g}}\gtrsim 0.9$~TeV (for $m_{\tilde{g}}\ll
m_{\tilde{q}}$)\cite{Aad:2012fqa,Chatrchyan:2012jx}.

The outline for the rest of the paper is as follows.
In section \ref{so10-theory} we briefly describe an SO(10) GUT model which can yield realistic fermion masses and mixings compatible with $t$-$b$-$\tau$ YU. In Section \ref{theory} we present the parameter space that we scan over, and describe how the MSSM gaugino mass {relations} can be obtained at $M_{\rm GUT}$.
In Section \ref{constraintsSection} we summarize the scanning procedure and the
experimental constraints applied in our analysis. In section \ref{muneg}  we discuss how SUSY threshold corrections, which are necessary to obtain $t$-$b$-$\tau$ YU and  radiative electroweak symmetry breaking (REWSB), {determine the} CP-even Higgs boson mass. In section \ref{results} we compare {the} results from the two packages, Isajet and SuSpect. {In section \ref{ft} we briefly discuss the little hierarchy problem in the presence of $t$-$b$-$\tau $ YU, and our} conclusions are presented in Section \ref{conclusions}.

\section{SO(10) GUT with  $t$-$b$-$\tau$ Yukawa unification\label{so10-theory}}

One of the main motivations of SO(10) GUT, in addition to gauge
coupling unification, is matter unification.
The spinor representation of SO(10) unifies all matter fermions of a
given family in a single multiplet ($16_i$), which also contains the right handed neutrino ($\nu_R$) that helps to generate light neutrino masses via the
see-saw mechanism \cite{seesawI}. The right handed neutrino can also naturally account for the baryon asymmetry of the Universe
via leptogenesis \cite{Fukugita:1986hr}.
Another virtue of SO(10) is that, in principle, the two MSSM Higgs doublets can be accommodated in a single ten dimensional representation, which then yields the following Yukawa couplings
\begin{align}
Y_{ij}\ 16_i\, 16_j\, 10_{\rm H}.
\label{10-yukawa}
\end{align}
Here $i,j=1, 2, 3$ stand for family indices and the SO(10) indices have been omitted for simplicity.
For the third generation quarks and  leptons, the interaction  in Eq.(\ref{10-yukawa}) yields  the following Yukawa coupling unification condition  at the  GUT scale \cite{big-422}
\begin{align}
Y_t = Y_b = Y_{\tau} = Y_{\nu_{\tau}}. \label{f1}
\end{align}
 It is interesting to note that in the gravity mediation SUSY breaking scenario \cite{Chamseddine:1982jx}, $t$-$b$-$\tau$ {YU condition leads to} LHC testable sparticle spectrum \cite{bigger-422,Gogoladze:2010fu} and even predicts a 125 GeV light CP-even Higgs boson mass \cite{Gogoladze:2011aa}.
 On the other hand, it is well known that the interaction in Eq.(\ref{10-yukawa}) leads
to a naive SO(10) relation: $N=U\propto D=L$, where $U$, $D$, $N$ and $L$ denote the Dirac mass matrices for up and down quarks, neutrinos and charge leptons respectively.  $U\propto D$ would imply vanishing quark flavor (Cabibbo-Kobayashi-Maskawa (CKM)) mixing   \cite{Langacker:1980js}, and $m_c^0/m_t^0=m_s^0/m_b^0$ which significantly  contradicts with experimental observations. The superscript zero refers to the parameters evaluated at the $M_{GUT}$. $D=L^T$,  which is a  naive SU(5) relation, would imply $m_s^0=m_{\mu}^0$ and $m_d^0=m_e^0$  and this relation also strongly disagrees with the measurements. So, it is obvious the interaction in Eq.(\ref{10-yukawa}) {should} be modified in order to accommodate the observed pattern of quarks and mixing.

 There are  two main approaches to avoid these problems and obtain realistic fermion masses and mixings. One way is to extend the Higgs sector and assume that the SM Higgs doublet fields are  superposition of fields {from} the  different {SO(10)} representations \cite{Babu:1992ia}. Another way is to introduce additional vector-like matter multiplets at the GUT scale \cite{Witten:1979nr} which mix in a nontrivial way with fermions in the 16 dimensional representation. It is equivalent to introducing non-renormalizable couplings which involve a non-singlet SO(10) field that develops a VEV  \cite{Anderson:1993fe}.
 Both these cases, however, allow  $t$-$b$-$\tau$ YU to be completely   destroyed or partially destroyed to yield $b$-$\tau$ YU.
  Therefore, in  order to maintain $t$-$b$-$\tau$ YU the following two conditions should be satisfied:
\begin{itemize}
\item Third generation {charged} fermions obtain their masses only from Eq.(\ref{10-yukawa}).
\item The MSSM {Higgs doublets} $H_u$ and $H_d$ reside solely in $10_H$.
\end{itemize}
 There {exist} models where
  $t$-$b$-$\tau$ YU is maintained {to} a very good approximation with realistic fermion masses and mixings \cite{Joshipura:2012sr}.

We consider the case {in which} the MSSM Higgs fields {are} contained in the $10_H$ representation and assume
 non-renormalizable couplings in the superpotential
\begin{align}
Y_{ij}\ 16_i\, 16_j\, 10 + Y^{\prime}_{ij}\, 16_i\,16_j\,\left(\frac{45}{M}\right)^n 10+ f_{ij}\frac{16_i\,16_j\, \overline{16}_H\,\overline{16}_H}{M}.
\label{sp1}
\end{align}
Here $M$,  a scale associated with the effective non-renormalizable interaction, could plausibly lie somewhere between the unification scale and {the reduced Planck mass ($M_{P}=2.4\times10^{18}$ GeV)}. In Eq.(\ref{sp1})  $n$ is an integer  and   
$45$ is an adjoint representation  of {SO(10)}.  We can have several $45$-dimensional fields with {VEVs in} different directions in the space spanned by the 45 generator of SO(10), as long as the SM gauge group is unbroken.
Ref. \cite{Anderson:1993fe} shows that in order to have the correct  naive SO(10) and SU(5) relationships for the first two families we  need to have at least  two  45 dimensional fields with VEVs in different directions in the SO(10) space.  If one of the 45-plet develops a VEV along the $B-L$ direction it can naturally lead to the so-called Georgi-Jarlskog relation \cite{Georgi:1979df}, $m_s^0\cong m_{\mu}^0/3$ and $m_d^0 \cong 3 m_e^0$. This will give correct masses after evolution of the RGEs to low scale.  The second 45-plet pointing towards $I_{3R}$, the third component of the right handed isospin group, will break $U\propto D$ and will allow for the correct CKM mixing matrix.

The pair of spinors $16_H+\overline{16}_H$ break the rank of the group from five to four and provides a mass for the right handed neutrinos. {The} third term in Eq.(\ref{sp1}) is responsible for generating the right handed neutrino masses. The adjoint 45-plet completes the breaking of SO(10) to the SM gauge group.  In order to keep the MSSM Higgs field in $10_H$, the term $h_i\,16_H\,16_H\, 10$ {allowed in principle}, should not be in the superpotential.
We can just assume the $h_i$ coupling to be  zero, or it can be forbidden by introducing an  additional symmetry.
  It was shown  in \cite{Anderson:1993fe} that the superpotential in Eq.(\ref{sp1}) perfectly describes {the observed} fermion masses and {mixings}.

 Having the MSSM Higgs fields in the $10_H$ representation, the doublet-triplet splitting problem can be solved in SO(10) using the missing VEV mechanism \cite{Dimopoulos:1981xm}. A variety of realistic models based on this mechanism have been constructed \cite{babubarr2}.

\section{Fundamental Parameter Space \label{theory}}

 It has been pointed out \cite{Martin:2009ad} that non-universal MSSM gaugino masses at $ M_{\rm GUT} $ can arise from non-singlet F-terms, compatible with the underlying GUT symmetry such as SU(5) and SO(10). The SSB
gaugino masses in supergravity  \cite{Chamseddine:1982jx} can arise, say, from the following
dimension five operator:
\begin{align}
 -\frac{F^{ab}}{2 M_{\rm
P}} \lambda^a \lambda^b + {\rm c.c.}
\end{align}
 Here $\lambda^a$ is the two-component gaugino field, $ F^{ab} $ denotes the F-component of the field which breaks SUSY, the indices $a,b$ run over
the adjoint representation {of the gauge group}. The resulting gaugino
mass matrix is $\langle F^{ab} \rangle/M_{\rm P}$, where the
supersymmetry breaking  parameter $\langle F^{ab} \rangle$
transforms as a singlet under the MSSM gauge group $SU(3)_{c}
\times SU(2)_L \times U(1)_Y$. The $F^{ab}$ fields belong to an
irreducible representation in the symmetric part of the direct product of the
adjoint representation of the unified group. This is a supersymmetric generalization of operators considered a long time ago \cite{Hill:1983xh}.

In SO(10), for example,
\begin{align}
({ 45} \times { 45} )_S = { 1} + { 54} + { 210} +
{ 770}.
\end{align}
If  $F$  transforms as a 54 or 210 dimensional
representation of SO(10) \cite{Martin:2009ad}, one obtains the following relation
among the MSSM gaugino masses at $ M_{\rm GUT} $ :
\begin{align}
M_3: M_2:M_1= 2:-3:-1 ,
\label{gaugino10}
\end{align}
where $M_1, M_2, M_3$ denote the gaugino masses of $U(1)$, $SU(2)_L$ and $SU(3)_c$
respectively. The low energy implications of this relation have recently been investigated in \cite{Okada:2011wd} without imposing YU. {In this paper} we consider the case with $ \mu > 0 $ and non-universal gaugino masses defined in Eq.(\ref{gaugino10}). In order to obtain the correct sign for the desired contribution to $ (g-2)_{\mu} $, we {set} $ M_{1} > 0 $, $ M_{2} > 0 $ and $ M_{3} < 0 $. Somewhat to our surprise, we find that this class of $ t$-$b$-$\tau $ YU models make a rather sharp prediction for {the lightest SM-like Higgs boson mass \cite{Gogoladze:2011aa}}. In addition, lower mass bounds on the masses of the squarks and gluino are obtained.

{Notice that in  general, {if} $F^{ab}$ transforms {non trivially} under {SO(10)}, the SSB terms such as the trilinear couplings and scalar mass terms are not necessarily universal at $M_{GUT}$}. However, we can assume, consistent with SO(10) gauge symmetry, that the coefficients associated with terms that violate the SO(10)-invariant form are suitably small, except for the gaugino term in Eq.(\ref{gaugino10}). {We also assume that D-term contributions to the SSB {terms} are much smaller compared with contributions from fields with non-zero auxiliary F-terms.}

Employing the boundary condition from Eq.(\ref{gaugino10}), one  can define the MSSM gaugino masses at $ M_{\rm GUT} $ in terms of the mass parameter $M_{1/2}$ :
\begin{align}
M_1= M_{1/2} \nonumber \\
M_2= 3M_{1/2} \nonumber \\
M_3= - 2 M_{1/2}.
 \label{gaugino11}
\end{align}
Note that $M_2$ and $M_3$ have opposite signs which, as we {will} show, is important {in} implementing  Yukawa coupling unification to a high accuracy. In order to quantify Yukawa coupling unification, we define the quantity $R_{tb\tau}$ as,
\begin{align}
R_{tb\tau}=\frac{ {\rm max}(y_t,y_b,y_{\tau})} { {\rm min} (y_t,y_b,y_{\tau})}.
\end{align}

We have performed random scans for the following parameter range:
\begin{align}
0\leq  m_{16}  \leq 10\, \rm{TeV} \nonumber \\
0\leq   m_{10} \leq 10\, \rm{TeV} \nonumber \\
0 \leq M_{1/2}  \leq 5 \, \rm{TeV} \nonumber \\
35\leq \tan\beta \leq 55 \nonumber \\
-3\leq A_{0}/m_{16} \leq 3
 \label{parameterRange}
\end{align}
 Here $ m_{16} $ is the universal SSB mass for MSSM sfermions, $ m_{10} $ is the universal SSB mass term for up and down MSSM Higgs masses, $ M_{1/2} $ is the gaugino mass parameter, $ \tan\beta $ is the ratio of the vacuum expectation values (VEVs) of the two MSSM Higgs doublets, $ A_{0} $ is the universal SSB trilinear scalar interaction (with corresponding Yukawa coupling factored out).  We use {the} central value $m_t = 173.1\, {\rm GeV}$ and 1$\sigma$ deviation    ($m_t = 174.2\, {\rm GeV}$) for top quark  in our analysis   \cite{:1900yx}. We choose a +1$\sigma$ deviation in $m_t$ since it leads to an increase in the Higgs mass and improves the prediction of the Higgs mass in our analysis. Our results however are not
too sensitive to one or two sigma variation in the value of $m_t$  \cite{Gogoladze:2011db}.
We use $m_b(m_Z)=2.83$ GeV which is hard-coded into Isajet.

\section{Constraints and Scanning Procedure\label{constraintsSection}}

We employ Isajet~7.84 \cite{ISAJET} and SuSpect 2.41 \cite{Djouadi:2002ze} interfaced with Micromegas 2.4 \cite{Belanger:2008sj} to perform random
scans over the fundamental parameter space. Isajet and SuSpect employ full two loop RGEs for the SSB parameters between $M_Z$ and $M_{GUT}$. {The} approach employed by both is similar, but there are some important differences that have been {previously studied} \cite{Allanach:2002pz, Baer:2005pv}. SuSpect assumes that the full set of MSSM RGEs are valid between $M_Z$ and $M_{GUT}$ and uses what is referred to as the `common scale approach' in \cite{Baer:2005pv}. In this approach the $\overline{DR}$ parameters are extracted at a common scale $M_{EWSB}=\sqrt{m_{{\tilde t}_1}m_{{\tilde t}_2}}$. Isajet, on the other hand, uses the `step-beta function approach' which means that it employs one-loop step beta functions for gauge and Yukawa couplings. These two approaches yield very similar results for most of the SUSY parameter space. There are some regions where the discrepancies get magnified \cite{Allanach:2002pz} and which we observe in our results as well. These differences will  be discussed in section \ref{results}.

{An} approximate error {of} around  $2$ GeV in the estimate of the
Higgs mass in Isajet and SuSpect largely arise from theoretical  uncertainties  in the calculation of the minimum of the scalar potential,  and
to a lesser extent from experimental uncertainties in the values
for $m_t$ and $\alpha_s$.

An important constraint comes from limits on the cosmological abundance of stable charged
particles  \cite{Nakamura:2010zzi}. This excludes regions in the parameter space
where  charged SUSY particles  become
the lightest supersymmetric particle (LSP). We accept only those
solutions for which one of the neutralinos is the LSP and saturates
the WMAP  bound on {the} relic dark matter abundance.

Micromegas is interfaced with Isajet and SuSpect to calculate the relic density and branching ratios $BR(B_s \rightarrow \mu^+ \mu^-)$ and $BR(b \rightarrow s \gamma)$. With these codes we implement the following random scanning procedure: A uniform and logarithmic distribution of random points is first generated in the parameter space given in Eq. (\ref{parameterRange}).
The function RNORMX \cite{Leva} is then employed
to generate a gaussian distribution around each point in the parameter space.  The data points
collected all satisfy
the requirement of radiative electroweak symmetry breaking  (REWSB),
with the neutralino in each case being the LSP. After collecting the data, we impose
the mass bounds on all the particles \cite{Nakamura:2010zzi} and use the
IsaTools package~\cite{Baer:2002fv}
to implement the various phenomenological constraints. We successively apply the following experimental constraints on the data that
we acquire from SuSpect and Isajet:

\begin{table}[h]\centering
\begin{tabular}{rlc}
$ 0.8 \times 10^{-9} \leq BR(B_s \rightarrow \mu^+ \mu^-) $&$ \leq\, 6.2 \times 10^{-9} \;
 (2\sigma)$        &   \cite{:2007kv}      \\
$2.99 \times 10^{-4} \leq BR(b \rightarrow s \gamma) $&$ \leq\, 3.87 \times 10^{-4} \;
 (2\sigma)$ &   \cite{Barberio:2008fa}  \\
$0.15 \leq \frac{BR(B_u\rightarrow
\tau \nu_{\tau})_{\rm MSSM}}{BR(B_u\rightarrow \tau \nu_{\tau})_{\rm SM}}$&$ \leq\, 2.41 \;
(3\sigma)$ &   \cite{Barberio:2008fa}  \\
 $ 0 \leq \Delta(g-2)_{\mu}/2 $ & $ \leq 55.6 \times 10^{-10} $ & \cite{Bennett:2006fi}
\end{tabular}\label{table}
\end{table}

%%%%%%%%%%%%%%%%%%%%%%%%%%%%%%%%%%%%%%%%%%%%%%%%%%%%%%%%%%%
%trim=l b r t
%\newpage
\begin{figure}[]
\newpage
%\vspace{-1cm}
\centering
\subfigure[Isajet 7.84]{\label{fig:1-a}{\includegraphics[scale=0.4]{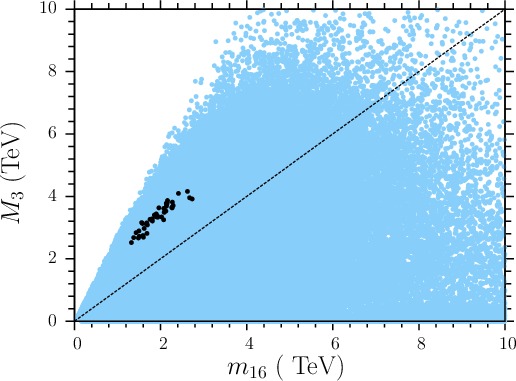}}}\hfill
\subfigure[SuSpect 2.41]{\label{fig:1-b}{\includegraphics[scale=0.4]{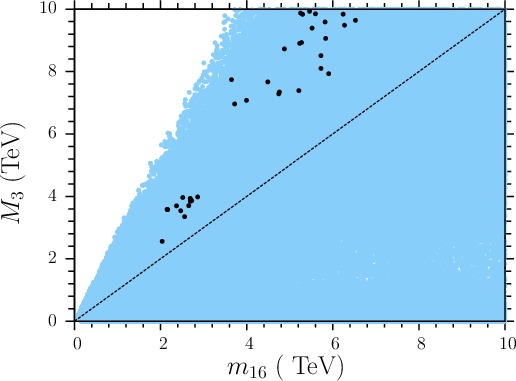}}}
\caption{Plot in the $M_3 - m_{16}$ planes. The panel on the left shows data points collected with Isajet 7.84 whereas the panel on the right shows data from SuSpect 2.41. The light blue points are consistent with REWSB and neutralino LSP. For both the panels, black points are subset of the light blue points and satisfy $R_{tb\tau} \le 1.03$ in panel \ref{fig:1-a}  and $R_{tb\tau} \le 1.05$ in panel \ref{fig:1-b}.
{The unit line is to guide the eye}.}
\label{fig:m3-m16}
\end{figure}

%%%%%%%%%%%%%%%%%%%%%%%%%%%%%%%%%%%%%%%%%%%%%%\vspace*{2mm}

%%%%%%%%%%%%%%%%%%%%%%%%%%%%%%%%%%%%%%%%%%%%%%%%%%%%%%%%%%%
%trim=l b r t
%\newpage
\begin{figure}[]
\newpage
%\vspace{-1cm}
\centering
{\label{fig:4-b}{\includegraphics[scale=0.4]{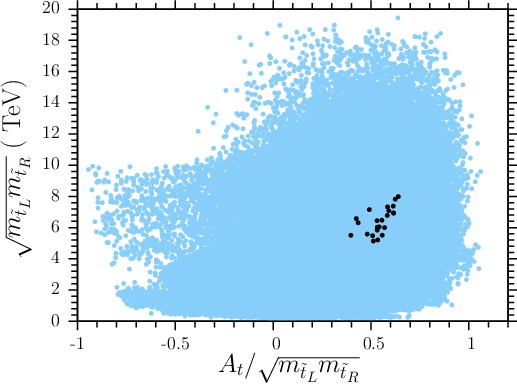}}}
\caption{Plot in the $\sqrt{m_{\tilde{t}_L} m_{\tilde{t}_R}} - A_t/ \sqrt{m_{\tilde{t}_L} m_{\tilde{t}_R}}$ planes. The data points shown are collected using Isajet. Color coding is the same as in Figure \ref{fig:m3-m16}.}
\label{fig: ms2}
\end{figure}

%%%%%%%%%%%%%%%%%%%%%%%%%%%%%%%%%%%%%%%%%%%%%%\vspace*{2mm}

%%%%%%%%%%%%%%%%%%%%%%%%%%%%%%%%%%%%%%%%%%%%%%%%%%%%%%%%%%%
%trim=l b r t
%\newpage
\begin{figure}[]
\newpage
%\vspace{-1cm}
\centering
\subfigure[Isajet 7.84]{\label{fig:3-a}{\includegraphics[scale=0.4]{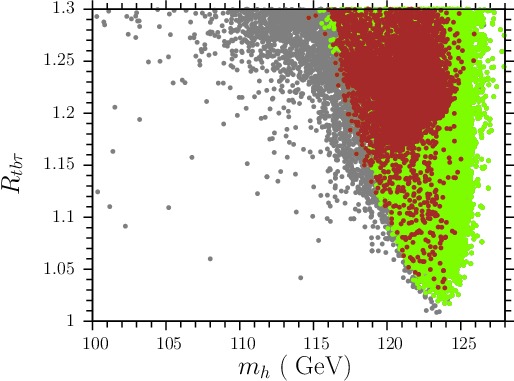}}}\hfill
\subfigure[SuSpect 2.41]{\label{fig:3-b}{\includegraphics[scale=0.4]{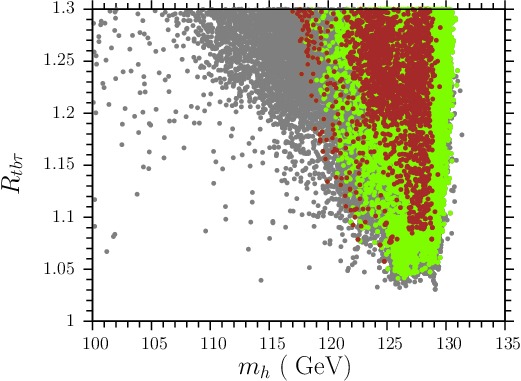}}}\\
\subfigure[Isajet 7.84 + SuSpect 2.41]{\label{fig:3-c}{\includegraphics[scale=0.4]{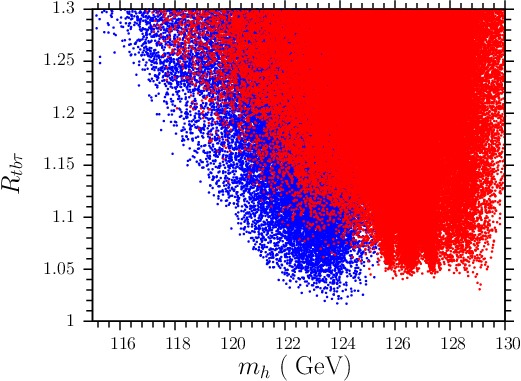}}}
\caption{Plot in the $R_{tb\tau} - m_h$ plane. Panel \ref{fig:3-a} shows results obtained from Isajet, while \ref{fig:3-b} shows results from obtained SuSpect. Gray points in \ref{fig:3-a} and \ref{fig:3-b} are consistent with REWSB and LSP neutralino. Green points form a subset of the gray and satisfy sparticle mass \cite{Nakamura:2010zzi} and B-physics constraints described in Section \ref{constraintsSection}.
 { In addition, we require that green points do no worse than the SM in terms of $ (g-2)_{\mu} $.}
 Brown points form a subset of the green points and satisfy $\Omega h^2 \le 1$. Panel \ref{fig:3-c} shows data collected with Isajet (blue) and SuSpect (red). The red and blue points satisfy the constraints imposed on the green points in panels \ref{fig:3-a} and \ref{fig:3-b}.}
\label{fig:R-mh}
\end{figure}

%%%%%%%%%%%%%%%%%%%%%%%%%%%%%%%%%%%%%%%%%%%%%%\vspace*{2mm}

\section{Higgs Mass Prediction\label{muneg}}

In this section we discuss the implicit relationship between the SUSY threshold corrections  to the Yukawa  {couplings},  REWSB and the light CP-even  Higgs mass. We begin with the SUSY
threshold corrections to the top, bottom and tau Yukawa couplings which
play a crucial role in $t$-$b$-$\tau$ Yukawa coupling unification.
In general, the bottom Yukawa coupling $y_b$ can receive large
threshold corrections, while the threshold corrections to $y_t$ are
typically smaller~\cite{Pierce:1996zz}. The scale at which Yukawa coupling unification
occurs is identified with $M_{\rm GUT}$, which is the scale of gauge coupling unification.
Consider first the unification between  $y_t ( M_{\rm GUT})$   and  $y_{\tau}(M_{\rm
GUT})$. The SUSY correction to the
tau lepton mass  is given by $\delta m_{\tau}=v\cos\beta \delta y_{\tau}$.
For the large $\tan\beta$ values of interest here, there is sufficient
freedom in the choice of $\delta y_{\tau}$ to achieve $y_t\approx y_{\tau}$
at $M_{\rm GUT}$. This freedom stems from the fact that
$\cos\beta \simeq 1/\tan\beta$ for large $\tan\beta$, and so we may
choose an appropriate $\delta y_{\tau}$ and $\tan\beta$ to give us both
the correct $\tau$ lepton mass and $y_t \approx y_{\tau}$ at $M_{\rm GUT}$.
The SUSY contribution to $\delta y_b$ has to be carefully monitored
in order to achieve Yukawa coupling unification $y_t (M_{\rm
GUT})\approx y_{b}(M_{\rm GUT}) \approx y_{\tau}(M_{\rm GUT})$.

We choose the sign of  $\delta y_i\, (i=t,b,\tau)$ from the perspective of
evolving $y_i$ from $M_{\rm GUT}$ to $M_{\rm Z}$. With this choice,
$\delta y_b$ must receive a negative contribution
($-0.27 \lesssim \delta y_b/y_b \lesssim -0.15$) in order to realize
Yukawa coupling unification.
The leading contribution to $\delta y_b$ arises from
the finite gluino and chargino loop corrections,
and in our sign convention, it is approximately given by~\cite{Pierce:1996zz}

\begin{align}
\delta y_b^{\rm finite}\approx\frac{g_3^2}{12\pi^2}\frac{\mu m_{\tilde g}
\tan\beta}{m_{\tilde b}^2}+
                         \frac{y_t^2}{32\pi^2}\frac{\mu A_t \tan\beta}{m_{\tilde t}^2},
\label{finiteCorrectionsEq}
\end{align}
where $g_3$ is the strong gauge coupling, $m_{\tilde g}$ is
the gluino mass, $m_{\tilde b}$ ($m_{\tilde t}$) is the
heaviest sbottom (stop) {mass}, and $A_t$ is the
top trilinear SSB  coupling.
The logarithmic corrections to $y_b$ are positive, which
leaves the finite corrections to provide for the correct overall negative
$\delta y_b$ in order to realize YU.

For models with gaugino mass unification or same sign gauginos, the gluino contribution
(first term in Eq.(\ref{finiteCorrectionsEq})) is positive for $\mu>0$. Thus, the chargino
contribution (second term in Eq.(\ref{finiteCorrectionsEq})) must play an essential role
in providing the required negative contribution to $\delta y_b$. This can be
achieved \cite{Baer:2008jn, Gogoladze:2009ug} only for
\begin{align}
m_{16} \gg M_{1/2},~~  m_{16}\gtrsim 6\,{\rm TeV}~~ {\rm and}~~ A_0/m_{16}\sim -2.6.
\label{rel-so10}
\end{align}

One could lower the sparticle mass spectrum  by considering opposite sign gaugino SSB terms \cite{Gogoladze:2010fu}  which is allowed by the 4-2-2
model \cite{pati}.  In  SO(10) GUT,
non-universality of SSB
gaugino masses with opposite signs can be generated through various SO(10) non singlet representations responsible for
 SUSY breaking \cite{Martin:2009ad}.

In particular, for $M_3<0$, $M_2>0$ and $\mu>0$, the gluino contribution to $\delta y_b$
has the correct sign to obtain the required $b$-quark mass, and furthermore, it is not necessary to have very strong relations among SSB fundamental parameters.
Yukawa coupling unification in this case is achieved for
\begin{align}
m_{16} \gtrsim 300\, {\rm
GeV}~~  {\rm and}~~ M_{3}(M_2)\geqslant m_{16},~~ {\rm  as~ well~ as~ for} ~~M_{3}(M_2) \leqslant  m_{16},
\label{rel-422}
\end{align}
as opposed to the parameter space given in Eq. (\ref{rel-so10}). This enables us to simultaneously
satisfy the requirements of  $t$-$b$-$\tau$ YU,
neutralino dark matter abundance and constraints from $(g-2)_\mu$, as well as a variety of
other bounds. But for the above mentioned  cases  the relation $m_{H_d}^2>m_{H_u}^2$ was imposed at $M_{\rm GUT}$.
It is well known that REWSB cannot be realized if we require $m_{H_d}^2=m_{H_u}^2$  at $M_{\rm GUT}$ and, in addition, demand  exact $t$-$b$-$\tau$ YU {(For Quasi YU case \cite{quasi} one can even have  $m_{H_d}^2=m_{H_u}^2=m_{16}^2$  at $M_{\rm GUT}$)}. Let's briefly review REWSB condition in light of YU.

The REWSB  minimization condition at tree level requires that
\begin{eqnarray}
\mu^2=\frac{m_{H_d}^2 - m_{H_u}^2\tan^{2}\beta}{\tan^{2}\beta -1} - \frac{M^{2}_{Z}}{2} \approx -m^{2}_{H_{u}}- \frac{M^{2}_{Z}}{2}.
\label{rewsb1}
\end{eqnarray}
Here {the} approximate equality works well for  large values for $\tan\beta$, {which} is the case {for} $t$-$b$-$\tau$ YU. {Plugging the tree level expression of the CP-odd Higgs mass
$m_A^2= m_{H_d}^2 + m_{H_u}^2+ 2\mu^2$ in Eq. (\ref{rewsb1})}, we obtain {the relation}
\begin{eqnarray}
m_{H_d}^2 - m_{H_u}^2\gtrsim M^{2}_{Z}+m_A^2.
\label{rewsb2}
\end{eqnarray}
On the other hand, {a} semi-analytical expression {for} $m_{H_d}^2 - m_{H_u}^2$ at $M_Z$, in terms of GUT scale fundamental {parameters,} has the following form \cite{Hempfling:1994sa}
\begin{eqnarray}
m_{H_d}^2 - m_{H_u}^2 \approx -0.13 m_{16}^2-0.26 m^2_{1/2}-0.04m_{1/2}A_0-0.01A^2,
\label{rge22}
\end{eqnarray}
which implies that in order to satisfy  the  condition in Eq. (\ref{rewsb2}), we {should require}
\begin{eqnarray}
M_{1/2}> m_{16}.
\label{cond1}
\end{eqnarray}
This requirement clearly contradicts the $t$-$b$-$\tau$ YU condition obtained in Eq. (\ref{rel-so10}) {if} gaugino mass unification {is} assumed at $M_{\rm GUT}$. This is the reason why REWSB cannot occur {if} precise Yukawa coupling unification and $m_{H_d}^2 =m_{H_u}^2$ {conditions} are imposed {at} $M_{{\rm GUT}}$.  On the other hand, the condition from Eq. (\ref{cond1}) can be {consistent} with the condition presented in Eq. (\ref{rel-422}), with {non-universal gaugino masses} at  $M_{\rm GUT}$. {The overlap} of conditions from Eqs. (\ref{cond1}) and  (\ref{rel-422}) gives very characteristic {relations} among {$m_{16}$ } and gaugino masses.

We  present in Figure  \ref{fig:m3-m16} the results of the scan over the parameter  space listed in Eq.(\ref{parameterRange}) in $M_3 -m_{16}$ plane. The light blue points are consistent with REWSB and $\tilde{\chi}_1^0$ LSP. Points in black correspond to $3\%$ or better $t$-$b$-$\tau$ YU ($R_{tb\tau}\leq 1.03$) for Isajet data. {For} SuSpect data, {the} black points correspond to $5\%$ or better $t$-$b$-$\tau$ YU.
In Section \ref{results} we will discuss the factors that can result in a few percent difference in the YU obtained in Isajet and SuSpect.
 We see that  $3\%$ in Isajet's case ($5\%$ in SuSpect's case ) or better $t$-$b$-$\tau$ YU can occur {for} $M_3$ slightly heavier {than} $m_{16}$, and
$M_3>2$ TeV. On the other hand, $M_3$ significantly {affects the} low scale stop quark mass and {an approximate} semi-analytic expression for
$m_{\tilde {t}_R}^2$ for   $t$-$b$-$\tau$ YU case is as follows:
\begin{eqnarray}
m_{\tilde {t}_R}^2 \approx 0.27 m_{16}+ 5.3
 M_{3}^2 + 0.4 M_{2}^2 + ...
\label{mt1}
\end{eqnarray}
Using Eq. (\ref{mt1}) we can obtain a rough lower bound for stop quarks, {namely} $m_{\tilde {t}_R} \gtrsim 4$ TeV.

{Next} we discuss how the {findings} presented in Figure \ref{fig:m3-m16} can affect {the} CP-even Higgs boson mass calculation.
For the actual calculation, {both} Isajet and SuSpect {employ a} more elaborate calculation {procedure}. {We include the} one-loop contributions to the CP-even Higgs boson mass \cite{at}:
\begin{eqnarray}
\left[ m_{h}^{2}\right] _{MSSM} \approx M_{Z}^{2}\cos ^{2}2\beta \left( 1-\frac{3
}{8\pi ^{2}}\frac{m_{t}^{2}}{v^{2}}t\right)
+\frac{3}{4\pi ^{2}}\frac{m_{t}^{4}}{v^{2}}\left[ t+\frac{1}{2}X_{t} \right],  \label{mh}
\end{eqnarray}
where
\begin{eqnarray}
v=174.1 {\rm \ GeV}, \ \
t =\log \left(
\frac{M_{S}^{2}}{M_{t}^{2}}\right),~~~
X_{t} &=&\frac{2\widetilde{A}_{t}^{2}}{M_{S}^{2}}\left( 1-\frac{\widetilde{A}%
_{t}^{2}}{12M_{S}^{2}}\right). \label{A1}
\end{eqnarray}%
Also $\widetilde{A}_{t}=A_{t}-\mu \cot \beta $, where
$A_{t}$ denotes the stop left and stop right soft
mixing parameter and $M_{\rm S}= \sqrt{m_{{\tilde t}_L}m_{{\tilde t}_R}}$.
Note that one loop radiative corrections to the CP-even Higgs mass depend logarithmically on  the stop quark mass and linearly on $X_t$. {These} two {quantities essentially determine the} radiative corrections to the CP-even Higgs boson mass. Because of this, it is interesting to present the result from Figure \ref{fig:m3-m16} in terms of $X_t$ and $M_S$.

In Figure \ref{fig: ms2} we show the results in the $M_S - A_t/M_S$ plane. The color coding is the same as in Figure
\ref{fig:m3-m16}. From the $M_S - A_t/M_S$ plane, we see that the black points lie in the interval 5 TeV $<M_S<$ 9 TeV. This interval reduces further if we require better YU. This means that for good YU, following Eq.(\ref{mh}) the lightest CP even Higgs boson {should} be relatively {heavy} owing to the logarithmic dependence on $M_S$.

Since the growth or decay of a logarithmic function is slow, the logarithmic dependence of $m_h$ on $M_S$ nicely explains the shape of colored points  in Figure \ref{fig:3-a} and \ref{fig:3-b}. Panel \ref{fig:3-a} shows results obtained from Isajet, while \ref{fig:3-b} shows results obtained using Suspect. {The gray} points in Figs. \ref{fig:3-a} and \ref{fig:3-b} are consistent with REWSB and LSP neutralino. {The green} points form a subset of the gray {points} and satisfy {the} sparticle mass and B-physics constraints described in Section \ref{constraintsSection}.  In addition, we require that green points do no worse than the SM in terms of $ (g-2)_{\mu} $. The brown points form a subset of the green points and satisfy $\Omega h^2 \le 1$.

 An intriguing feature of Figures \ref{fig:3-a} and \ref{fig:3-b} is that the minima of the distribution  occurs {at the Higgs mass value} very close to the observed mass of the SM-like Higgs at the LHC. In other words, nearly perfect {$t$-$b$-$\tau $ YU} {prefers} the current favored value of $m_h$. We can also understand  from  Figure \ref{fig: ms2} why {the} minima in  the Figure \ref{fig:3-a} and \ref{fig:3-b} {have} relatively small {widths}.
In Figure \ref{fig: ms2}, from the $M_S - A_t/M_S$ plane, {we see} that the ratio $A_t/M_S$ lies in the very small interval  0.3$<A_t/M_S<$0.7.  On the other hand, it is known \cite{xt-mh-mass} that the CP even Higgs boson obtains  significant contributions from $A_t$ {if} $A_t/M_S > 1$. We can therefore conclude that there is no significant contribution from the finite corrections to the CP even Higgs boson mass {if} we have almost perfect YU, and the Higgs mass is mostly generated from the logarithmic corrections. This is why {the} minima in Figures \ref{fig:3-a} and \ref{fig:3-b} {are} not wide.

In Figures \ref{fig:3-c} we show the overlap of the data from Isajet 7.84 and SuSpect 2.41.
All the points shown are consistent with REWSB, LSP neutralino, {and} satisfy {the} sparticle mass and   B-physics constraints described in Section \ref{constraintsSection}.  The blue {(red)} points show results obtained from Isajet {(SuSpect)}. We can see that the minima of the data distribution from Isajet predicts {a} Higgs mass $m_h \sim 124$ GeV, whereas {SuSpect} predicts $m_h \sim 126$ GeV. {For lower} values of $R_{tb\tau}$, the results from {the} two packages overlap around $m_h \approx 125$ GeV. This observation  makes {the predicted value} of Higgs mass {close to} 125 GeV {obtained from $t$-$b$-$\tau $ YU condition quite reliable.}

\section{Higgs and Sparticle Spectroscopy From Isajet and SuSpect \label{results}}

In this section we compare the allowed parameter spaces and  the sparticle spectroscopy obtained from {Isajet and SuSpect}. {A} comparison of the uncertainties in the sparticle spectroscopy of different packages was done in \cite{Allanach:2002pz}, and, in particular, the threshold effects were compared in \cite{Baer:2005pv}. The approach used by the two programs is very similar in that both use two loop {RGE running for the} gauge and Yukawa {couplings}. But there are some factors that can lead to numerical differences of a few percent.
These include the scale at which the sparticles are integrated out of the theory. In Isajet {the} SSB parameters are extracted from RGE running at their respective mass scales $m_i=m_i(m_i)$, whereas in SuSpect these parameters are extracted at $M_{EWSB}(\equiv \sqrt{m_{{\tilde t}_1}m_{{\tilde t}_2}})$.
   SuSpect uses $\alpha_s$ in the $\mathrm{\overline{ DR}}$ scheme, whereas Isajet uses the $\mathrm{\overline{ MS}}$ value.
 Another source of discrepancy can be the use of bottom pole mass by SuSpect $m_b= m_b(M_{Z})$, whereas Isajet uses the mass at the SUSY scale $m_b= m_b(M_{SUSY})$.
 The default guess of the sparticle masses at the beginning of the RG evolution process is different in Isajet and SuSpect.
 SuSpect assumes that the full set of MSSM RGEs are valid between $M_Z$ and $M_{GUT}$. This is also true for Isajet, except that it employs one-loop step-beta functions for gauge and Yukawa couplings.
 %Loop corrections in the minimization of the scalar potential, which are effected by the whole SUSY spectrum. To get the SUSY spectrum a loop corrected value of $\mu$ has to be guessed. The manner in which this value is guessed by the two codes can lead to a difference.

%\newpage

%%%%%%%%%%%%%%%%%%%%%%%%%%%%%%%%%%%%%%%%%%%%%%%%%%%%%%%%%%%
%trim=l b r t
%\newpage
\begin{figure}[]
\newpage
\vspace{-1cm}
\centering
{\label{fig:isa-a}{\includegraphics[scale=0.4]{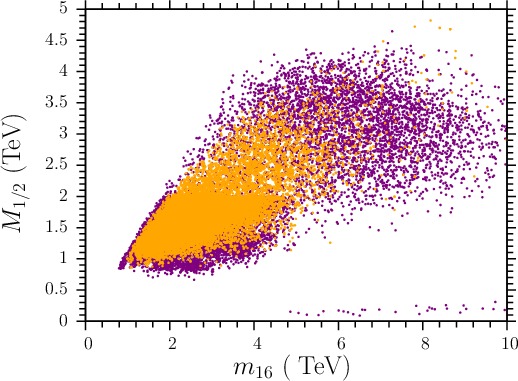}}}\hfill
{\label{fig:isa-b}{\includegraphics[scale=0.4]{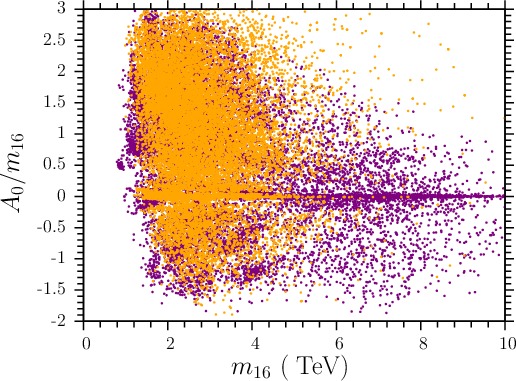}}}\\
\vspace{0.3cm}
{\label{fig:isa-c}{\includegraphics[scale=0.4]{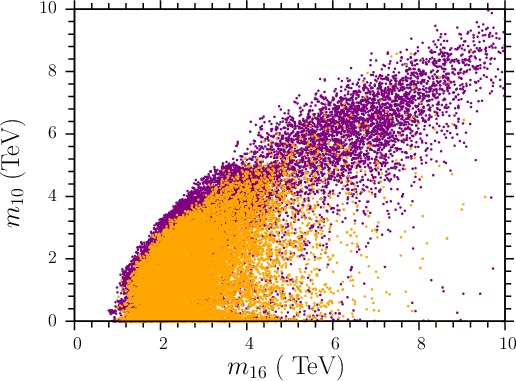}}}\hfill
{\label{fig:isa-b}{\includegraphics[scale=0.4]{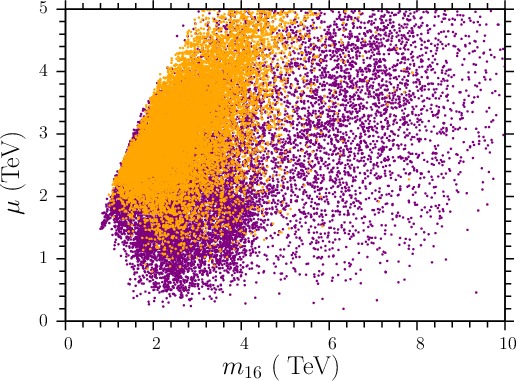}}}\\
\caption{Plots in the $M_{1/2} - m_{16}$,  $A_0/m_{16} - m_{16}$, $m_{10} - m_{16}$ and $\mu - m_{16}$  planes. All the points shown are consistent with REWSB, LSP neutralino and satisfy sparticle mass \cite{Nakamura:2010zzi} and B-
physics constraints described in Section \ref{constraintsSection}.
 We also require that the points do no worse than the SM in terms of $ (g-2)_{\mu} $.
In addition the satisfy the Higgs mass range $122  \mathrm{\ GeV} < m_h  < 128 \mathrm{\ GeV}$ and $R_{tb\tau} < 1.2$. The purple points show results obtained from SuSpect and yellow points is the data collected using Isajet.}
\label{fig:isajet-SuSpect}
\end{figure}

%%%%%%%%%%%%%%%%%%%%%%%%%%%%%%%%%%%%%%%%%%%%%%\vspace*{2mm}

 In Figure \ref{fig:isajet-SuSpect} we show our results in the $M_{1/2} - m_{16}$,  $A_0/m_{16} - m_{16}$, $m_{10} - m_{16}$ and  $\mu - m_{16}$  planes.  All {of} the points shown are consistent with REWSB, LSP neutralino and satisfy sparticle mass and B-physics constraints described in Section \ref{constraintsSection}. In addition, the points shown satisfy the {condition} $122  \mathrm{\ GeV} < m_h  < 128 \mathrm{\ GeV}$ and $R_{tb\tau} < 1.2$. The purple points show results obtained from SuSpect, whereas {the} yellow points {correspond to} the data collected using Isajet.

We {observe} that there are some small but  notable differences in the results obtained from the two packages. It is interesting to note that for the regions where the points are {more dense} there is good overlap between the two packages. In  Figure \ref{fig:isajet-SuSpect} under the very dense yellow points {there also lie} dense purple points. From
$M_{1/2} - m_{16}$,  $A_0/m_{16} - m_{16}$, $m_{10} - m_{16}$ panels, we see that the best agreement between solutions obtained from  Isajet and SuSpect {occurs for} $m_{16}<6$ TeV, $M_{1/2}<3$ TeV and $m_{10}< 4$ TeV.

\begin{figure}[]
\newpage
%\vspace{-1cm}
\centering
{\label{fig:isa-d}{\includegraphics[scale=0.4]{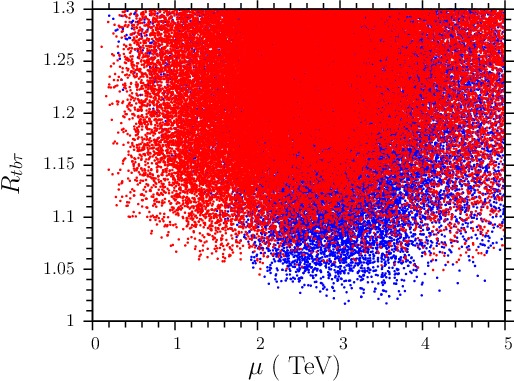}}}\hfill
{\label{fig:isa-e}{\includegraphics[scale=0.4]{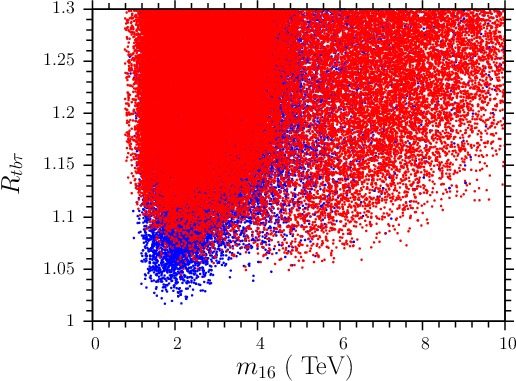}}}
\caption{Plots in the $R_{tb\tau} - \mu$, $R_{tb\tau} - m_{16}$  planes. Color coding is the same as in Figure \ref{fig:3-c}.}% except that the bounds on $m_h$ and $R_{tb\tau}$ have not been applied.}
\label{fig:isajet-SuSpect2}
\end{figure}

In the $\mu - m_{16}$ plane of Figure \ref{fig:isajet-SuSpect} we see that the solutions from SuSpect  with 20$\%$ or better $t$-$b$-$\tau$ YU
have lower values of $\mu$ {compared to the $\mu$ values} from Isajet. {To exemplify this further}, in Figure \ref{fig:isajet-SuSpect2} {we show a} plot in $R_{tb\tau} - \mu$ {plane}. We see that requiring $20\%$ or better $t$-$b$-$\tau$ YU allows solutions from SuSpect with $\mu\approx 200$ GeV, while requiring the same for Isajet {yields} $\mu \gtrsim 1$ TeV. On the other hand, requiring YU better then $5\%$ leads to similar limits on the values of $\mu$ from Isajet and SuSpect, i.e., $\mu \gtrsim 2$ TeV.
It was noted in \cite{Baer:2005ky} that the two codes can differ notably in {regions} with low $\mu$ and $\tan\beta$ {values}. This difference can stem from the factors {previously} mentioned, {and which may} have important implications for natural SUSY.

 We {observe} from Figure  \ref{fig:isajet-SuSpect2} that SuSpect does not yield YU better than 5\%, {with a} minimum value of $R_{tb\tau} \sim 1.05$. Isajet, however, predicts {even} better YU  with the minimum value of $R_{tb\tau} \sim 1.02$.
 {This} few percent difference  {could} be {due to} the way threshold effects are evaluated by the two codes \cite{Baer:2005pv}.
 The $R_{tb\tau}  -  m_{16}$ plane shows that in order to have YU better than $5\%$ the results from Isajet require $600$ GeV $<m_{16}<$ 2.5 TeV.

%%%%%%%%%%%%%%%%%%%%%%%%%%%%%%%%%%%%%%%%%%%%%%%%%%%%%%%%%%%
%trim=l b r t
%\newpage

%%%%%%%%%%%%%%%%%%%%%%%%%%%%%%%%%%%%%%%%%%%%%%%%%%%%%%%%%%%
%trim=l b r t

\begin{figure}[]
\begin{center}
\includegraphics[scale=0.4]{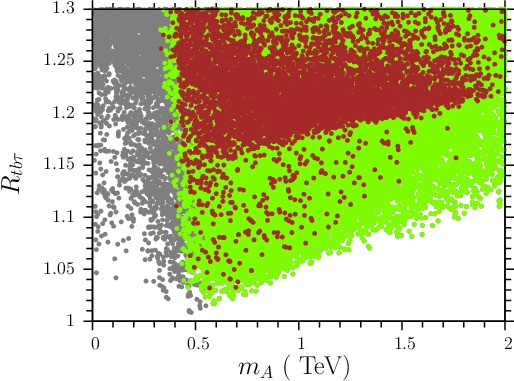}
\includegraphics[scale=0.4]{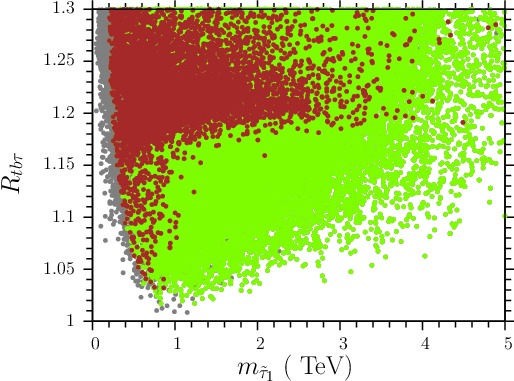}\\
\vspace{0.3cm}
\includegraphics[scale=0.4]{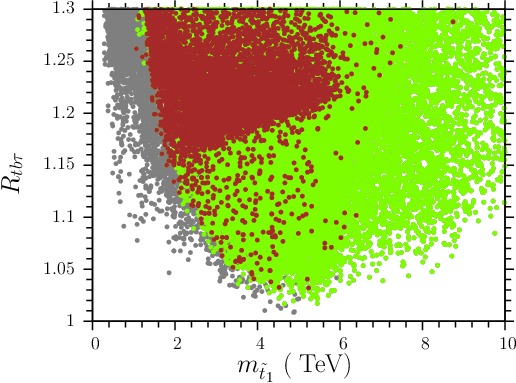}
\includegraphics[scale=0.4]{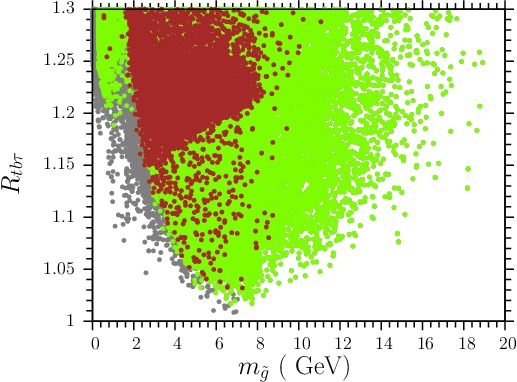}
\vspace{0.3cm}
\includegraphics[scale=0.4]{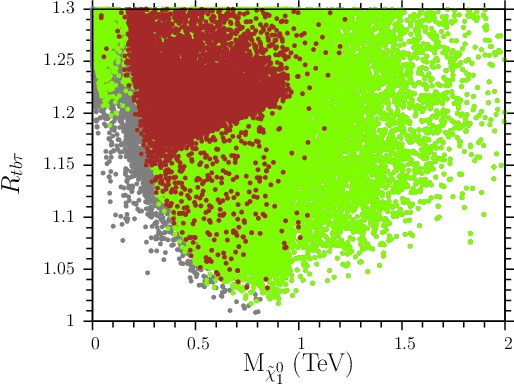}
\includegraphics[scale=0.4]{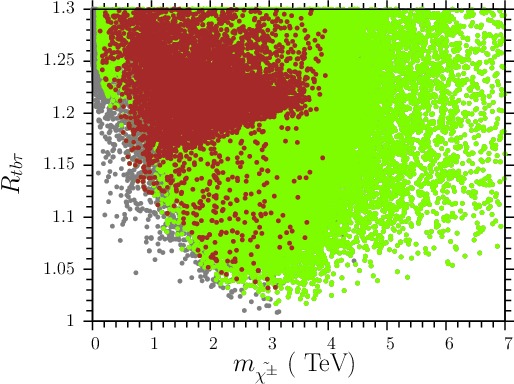}
\end{center}
\caption{Plot in the $R_{tb\tau} - m_A$, $R_{tb\tau} - m_{\tilde{\tau}_1}$,   $R_{tb\tau} - m_{\tilde{t}_1}$,
$R_{tb\tau} - m_{tilde{g}}$,  $R_{tb\tau} - m_{\tilde{\chi}^0_1}$ and  $R_{tb\tau} - m_{\tilde{\chi}^{\pm}}$ planes. The color coding is the same as in Figure \ref{fig:R-mh}.}
\label{fig:spectrum}
\end{figure}

%%%%%%%%%%%%%%%%%%%%%%%%%%%%%%%%%%%%%%%%%%%%%%\vspace

%%%%%%%%%%%%%%%%%%%%%%%%%%%%%%%%%%%%%%%%%%%%%%%%%%%%%%%%%%%
%trim=l b r t
%\newpage
\begin{figure}[]
\newpage
\vspace{-1cm}
\centering
{{\includegraphics[scale=0.4]{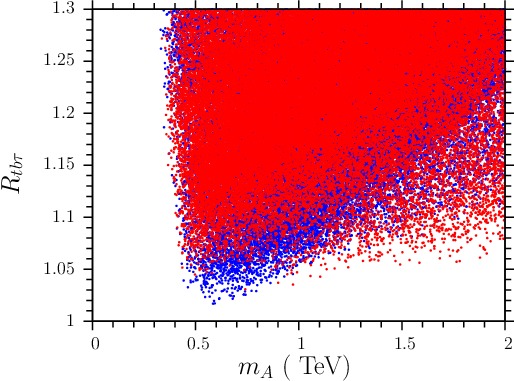}}}\hfill
%{{\includegraphics[scale=0.4]{plot-ratio-mtab5.jpg}}}\\
\caption{Plot in the   $R_{tb\tau} - m_A$ planes. {Color} coding is the same as in Figure \ref{fig:3-c}.}
\label{fig:maa22}
\end{figure}

%%%%%%%%%%%%%%%%%%%%%%%%%%%%%%%%%%%%%%%%%%%%%%\vspace*{2mm}

In Figure \ref{fig:spectrum} we show results in the $R_{tb\tau} - m_A$, $R_{tb\tau} - m_{\tilde{\tau}_1}$,   $R_{tb\tau} - m_{\tilde{t}_1}$,
$R_{tb\tau} - m_{\tilde{g}}$,  $R_{tb\tau} - m_{\tilde{\chi}^0_1}$ and  $R_{tb\tau} - m_{\tilde{\chi}^{\pm}}$ planes. The color coding is the same as in Figure  \ref{fig:R-mh}. The data collected with Isajet is used to make the plots in this figure. All panels in Figure  \ref{fig:spectrum} indicate that the model predicts relatively narrow ranges for the sparticle masses corresponding to the best $t$-$b$-$\tau$ YU. {The} sparticles are heavy enough to evade observation at current LHC energies, but a signal would hopefully be observed during the 14 TeV LHC run.

From the $R_{tb\tau} - m_A$ plane we {see} that just from the REWSB condition, $t$-$b$-$\tau$ YU {better than $5\%$} predicts that we cannot have a very {light CP-odd Higgs boson ($A$)}.
 {Similar} bounds apply for the {heavier} CP even $H$ {boson} and charged ${H^{\pm}}$ {bosons}, since, in the so-called decoupling limit, $m_A^2\gg M_Z^2$, we have $m_H^2 \simeq m_A^2$ and $m_{H^{\pm}}\simeq m_A^2+ M^2_W$ \cite{Martin:1997ns}, {where} $M_W$ stands for the $W$-boson mass. The  leading  decay modes for the heavy Higgs bosons are $H, A \rightarrow b\, \bar{b}$
 and $H, A \rightarrow \tau^+ \tau^-$.  The heavy Higgs production cross section and branching {ratios} depends on $\tan\beta$  and {its} mass. {The} MSSM sparticle dependence  appears {through} gluino-squark, Higgsino-squark, bino-sfermion and wino-sfermion {loops}.
The bound on $m_A$ in our scenario is more relaxed since we have non universal gauginos  with opposite signs for $M_2$ and $M_3$ at the GUT scale. We also {find} that good $t$-$b$-$\tau$ YU {requires} $M_2/M_3>1.5$. This will {alter the} gluino-squark and wino-sfermion loop {contributions} to the heavy Higgs production cross section and branching {ratios} \cite{Altmannshofer:2012ks}.

Applying all the collider and B-physics constraints we obtain a lower bound for $m_A$ which is very close to the value corresponding to best {YU}. Restricting to $5\%$ or better unification and including {the} constraints presented in Section \ref{constraintsSection}, we obtain the bound 400 GeV$\lesssim M_A \lesssim$  1 TeV. {The lower} bound is very close to the current experimental limit \cite{Chatrchyan:2012vp} obtained from the GUT scale gaugino unification condition and can be further tested in near future.

 In the  $R_{tb\tau} - m_{\tilde{\tau}_1}$ plane we can observe that the {preferred} values for the stau lepton mass from the point of view of good YU ($R_{tb\tau}< 1.05$) is in the interval 500 GeV $ \lesssim m_{\tilde{\tau}_1} \lesssim $ 1.5 TeV. The search for a stau {in} this mass {range} is challenging at the LHC. In this model, {the} stau {is} the NLSP, {and this can yield} the correct relic abundance through neutralino-stau coannihilation.

{The} lightest {colored sparticle} in this scenario is {one of the stops}. The {preferred} mass as {seen} from the  $R_{tb\tau} - m_{\tilde{t}_1}$ plane,
  is  around 3-4 TeV. For the gluino, the mass according to the $R_{tb\tau} - m_{\tilde{g}}$ plane is around 5-6 TeV. {In principle} {they} can be {found} at the LHC.
 We can also see from Figure \ref{fig:spectrum} that the lightest neutralino, for $R_{tb\tau}< 1.05$, is around 500 GeV, and {the preferred value} from the point of view YU is $ \sim $ 700 GeV. The model also {predicts the} charginos to be heavier than 2 TeV.

 Since {the pseudoscalar} $A$ boson {is being searched for} at the LHC, we present  in Figure \ref{fig:maa22}
 the combined {results} from Isajet and SuSpect. {The color} coding is the same as in Figure    \ref{fig:spectrum}, and we see that {the agreement between the two programs is quite satisfying}.

\begin{figure}[]
\begin{center}
\includegraphics[scale=0.4]{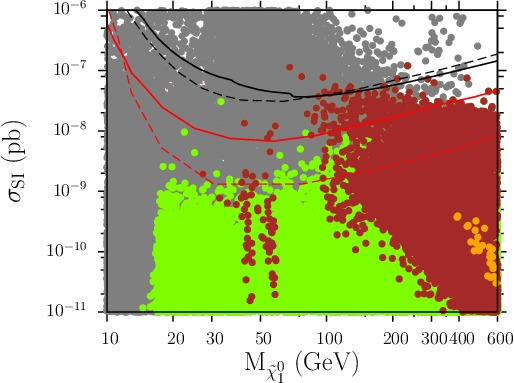}
\includegraphics[scale=0.4]{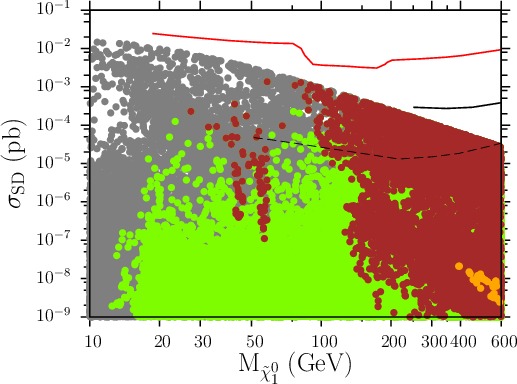}
\end{center}
\caption{Plots in the  $\sigma_{\rm SI} - m_{\tilde{\chi}_1^{0}}$ and $\sigma_{\rm SD} - m_{\tilde{\chi}_1^{0}}$
planes. The cross sections are calculated using Isajet. Points shown in gray are consistent with REWSB and LSP neutralino. Green points form a subset of the gray and satisfy sparticle mass \cite{Nakamura:2010zzi} and B-physics constraints described in Section \ref{constraintsSection}. In addition, we require that green points do no worse than the SM in terms of $ (g-2)_{\mu} $.
Brown points form a subset of the green points and satisfy $\Omega h^2 \le 1$. The orange points are a subset of the brown points and satisfy $R_{tb \tau} \le 1.1$
 In the $\sigma_{\rm SI}$ -
$m_{\tilde{\chi}_1^{0}}$ plane, the current and future bounds from the CDMS experiment are represented as black (solid and dashed) lines  and as red (solid and dotted) lines for the Xenon experiment. The right panel shows the $\sigma_{\rm SD}$ -
$m_{\tilde{\chi}_1^{0}}$ plane with the current bounds from Super
K (solid red line) and IceCube (solid black line) and future reach of
IceCube DeepCore (dotted black line).}
\label{fig:si-sd}
\end{figure}

 In Figure \ref{fig:si-sd} we show the implication of our analysis for direct detection of dark matter. Plots are shown in the  $\sigma_{\rm SI}  -  m_{\tilde{\chi}_1^{0}}$ and $\sigma_{\rm SD}  -  m_{\tilde{\chi}_1^{0}}$
planes. {The gray} points in the figure are consistent with REWSB and LSP neutralino. {The green} points form a subset of the gray {points} and satisfy {the} constraints described in Section \ref{constraintsSection}. {The brown} points form a subset of the green points and satisfy $\Omega h^2 \le 1$, and {the} orange points form a subset of the brown points with {10\% or better YU} ($R_{tb \tau} \le 1.1$). The left panel shows the current and future bounds from CDMS as black (solid and dashed) lines, {and} as red (solid and dotted) lines for the Xenon experiment. The right panel also shows the current bounds from Super
K (solid red line) and IceCube (solid black line), and future reach of
IceCube DeepCore (dotted black line).

 We can see that the parameter space of this model  representing {neutralino-stau} coannihilation can be tested with these experiments. However {models} with YU better than 10\%  yield tiny cross sections which are well below the sensitivity of these experiments. We {also} observe that good YU predicts a heavy neutralino $(m_{\tilde{\chi}_1^{0}} \gtrsim 400 \rm \ GeV)$. It is, however, interesting that these orange points also predict a 125 GeV Higgs mass {as seen} in {Figures} \ref{fig:3-a} and \ref{fig:3-b}. Therefore, $t$-$b$-$\tau $ YU not only predicts a 125 GeV Higgs but also a relatively heavy dark matter {LSP} which {coannhilates} with the stau to {yield} the correct relic abundance, and also yields tiny cross sections well below the sensitivity of current experiments.

{In Table \ref{tab1} we present four benchmark points with good YU and Higgs mass $\sim$ 125 GeV. The points shown also satisfy the constraints described in Section \ref{constraintsSection}. Points 1 and 2 represent solutions that yield the best YU in Isajet and SuSpect. As described earlier, Isajet {yields} YU {as good as} $\sim$ 2\%, whereas in SuSpect it is $\sim$ 5\%. Point 3 depicts stau coannihilation in addition to a 124 GeV Higgs and $R_{tb\tau}=1.03$. Point 4 shows that good YU can be attained with the sfermions and Higgs nearly degenerate at $M_{GUT}$, i.e., $m_{16}\simeq m_{10}$.}

%%%%%%%%%%%%%%%%%%%%%%%%%%%%%%%%%%%%%%%%%%%%%%%%%%%%%%%%%%%
%trim=l b r t

%%%%%%%%%%%%%%%%%%%%%%%%%%%%%%%%%%%%%%%%%%%%%%\vspace*{2mm}

\begin{figure}[]
\begin{center}
\includegraphics[scale=0.4]{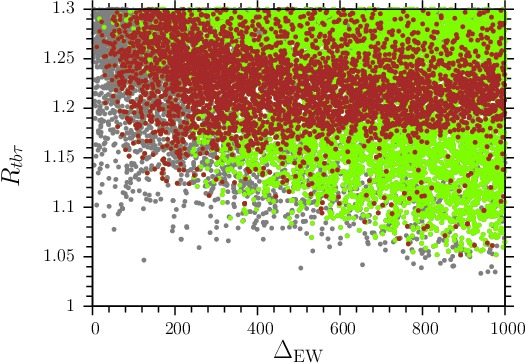}
\includegraphics[scale=0.4]{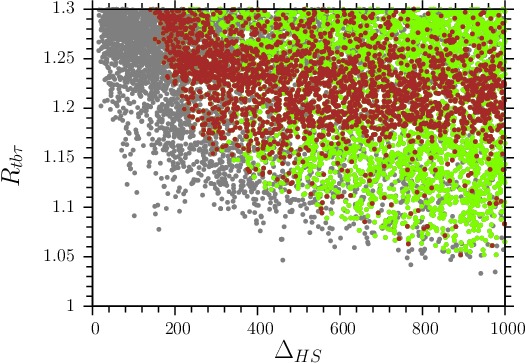}
\end{center}
\caption{Plots in the $R_{tb\tau}$-$\Delta_{EW}$ and $R_{tb\tau}$-$\Delta_{HS}$ planes. Color coding is the same as in Figure \ref{fig:R-mh}.}
\label{fig:ft22}
\end{figure}

\section{Fine tuning constraints for little hierarchy \label{ft}}

The latest (7.84) version of  ISAJET \cite{ISAJET} calculates the  fine-tuning conditions related to the little hierarchy problem at Electro Weak ($EW$) scale
and at the GUT scale ($HS$). We will briefly describe these parameters in this section.

After including the one-loop effective potential contributions to the tree level MSSM Higgs potential, {the Z} boson mass is given by the following relation:
\begin{equation}
\frac{M_Z^2}{2} =
\frac{(m_{H_d}^2+\Sigma_d^d)-(m_{H_u}^2+\Sigma_u^u)\tan^2\beta}{\tan^2\beta
-1} -\mu^2 \; .
\label{eq:mssmmu}
\end{equation}
The $\Sigma$'s stand for the contributions coming from the one-loop effective potential (For more details see ref. \cite{Baer:2012mv}). All parameters  in Eq. (\ref{eq:mssmmu}) are defined at $EW$ scale.

In order to measure the $EW$ scale fine-tuning condition associated with the little hierarchy problem, the following definitions are used \cite{Baer:2012mv}:
\begin{equation}
 C_{H_d}\equiv |m_{H_d}^2/(\tan^2\beta -1)|,\,\, C_{H_u}\equiv
|-m_{H_u}^2\tan^2\beta /(\tan^2\beta -1)|, \, \, C_\mu\equiv |-\mu^2 |,
\label{cc1}
\end{equation}
 with
each $C_{\Sigma_{u,d}^{u,d} (i)}$  less
than some characteristic value of order $M_Z^2$.
Here, $i$ labels the SM and supersymmetric
particles that contribute to the one-loop Higgs potential.
For the fine-tuning condition we have
\begin{equation}
 \Delta_{\rm EW}\equiv {\rm max}(C_i )/(M_Z^2/2).
\label{eq:ewft}
\end{equation}
Note that Eq. (\ref{eq:ewft}) defines the fine-tuning  condition  at $EW$ scale without addressing
the question of the origin of the parameters that are involved.

In most SUSY breaking scenarios the parameters in
Eq.~(\ref{eq:mssmmu}) are defined at a scale higher than $M_{EW}$.
In order to fully address  the fine-tuning condition we need to  check the relations
 among the parameters involved  in Eq.~(\ref{eq:mssmmu}) at high scale. We relate the parameters at low and high scales as follows:
 \begin{equation}
 m_{H_{u,d}}^2=
m_{H_{u,d}}^2(M_{HS}) +\delta m_{H_{u,d}}^2, \ \,\,\,
\mu^2=\mu^2(M_{HS})+\delta\mu^2.
\end{equation}
 Here
$m_{H_{u,d}}^2(M_{HS})$ and $\mu^2(M_{HS})$ are the corresponding
parameters renormalized at the high scale, and
$\delta m_{H_{u,d}}^2$, $\delta\mu^2$ measure how the given parameter is changed due to renormalization group evolution (RGE).
 Eq.~(\ref{eq:mssmmu}) can be re-expressed in the form
\begin{eqnarray}
\frac{m_Z^2}{2} &=& \frac{(m_{H_d}^2(M_{HS})+ \delta m_{H_d}^2 +
\Sigma_d^d)-
(m_{H_u}^2(M_{HS})+\delta m_{H_u}^2+\Sigma_u^u)\tan^2\beta}{\tan^2\beta -1}
\nonumber \\
&-& (\mu^2(M_{HS})+\delta\mu^2)\;.
\label{eq:FT}
\end{eqnarray}
Following ref. \cite{Baer:2012mv}, we  introduce the  parameters:
\begin{eqnarray}
&B_{H_d}\equiv|m_{H_d}^2(M_{HS})/(\tan^2\beta -1)|,
B_{\delta H_d}\equiv |\delta m_{H_d}^2/(\tan^2\beta -1)|, \nonumber \\
&B_{H_u}\equiv|-m_{H_u}^2(M_{HS})\tan^2\beta /(\tan^2\beta -1)|, B_{\mu}\equiv|\mu^2(M_{HS})|, \nonumber \\
&B_{\delta H_u}\equiv|-\delta m_{H_u}^2\tan^2\beta /(\tan^2\beta -1)|,
  B_{\delta \mu}\equiv |\delta \mu^2|,
  \label{bb1}
\end{eqnarray}
and  the
high scale fine-tuning measure $\Delta_{\rm HS}$ is defined to be
\begin{equation}
\Delta_{\rm HS}\equiv {\rm max}(B_i )/(M_Z^2/2).
\label{eq:hsft}
\end{equation}

The current experimental bound on the chargino mass ($m_{\widetilde W}> 103$ GeV) \cite{Nakamura:2010zzi} indicates that either $\Delta_{EW}$ or $\Delta_{HS}$ cannot be less than 1. The quantities $\Delta_{EW}$ and  $\Delta_{HS}$ measure the sensitivity of the Z-boson mass to the parameters defined in Eqs. (\ref{cc1}) and (\ref{bb1}), such that $(100/\Delta_{EW})\%$  ($(100/\Delta_{HS})\%$) is the degree of fine-tuning at the corresponding scale.

Based on the definition of high and low scale fine tuning described above we show results in the $R_{tb\tau}-\Delta_{EW}$
and $R_{tb\tau}-\Delta_{HS}$ {planes (using Isajet)} in Figure \ref{fig:ft22}. We see that the low scale little hierarchy problem becomes more severe for the $t$-$b$-$\tau$ YU case compared to what we have in the constrained MSSM (CMSSM), but high scale fine tuning is {at} the same level as in the CMSSM \cite{Gogoladze:2012yf}. For $t$-$b$-$\tau$ YU {of} around $5\%$, the $EW$ fine tuning parameter $\Delta_{EW} \sim 800$ and the $HS$ fine tuning parameter is also  $\Delta_{HS} \sim 800$. As mentioned above, the
fine tuning condition has to be scale invariant which means that cancellation between parameters  at a particular scale cannot  be  more severe compared to same conditions at another scale. Based on this assumption the little hierarchy problem in this model remains the same as we have when gaugino universality is assumed in the theory \cite{Baer:2012mv,Gogoladze:2012yf}.

%\newpage
%%%%%%%%%%%%%%%%%%%%%%%%%%%%%%%%%%%%%%%%%%%%%%\vspace*{2mm}
\begin{table}[p!]
\centering
%\begin{tabular}{|p{3cm}|p{3cm}|p{3cm}|p{3cm}|p{3cm}|}
\begin{tabular}{|c|c|c|c|c|}
\hline
\hline
                 	&	 Isajet  	&	 SuSpect  	&	 Isajet  	&	 Isajet  	\\
%\times 10^{									
\hline									
%\times 10^{- \times 10^{ \times 10^{									
									
$m_{10} $         	&$	          4.19\times 10^{2}	$&$	         3.82\times 10^{3}	$&$	          4.49\times 10^{2}	$&$	          1.94\times 10^{3}	$\\
$m_{16} $         	&$	          2.13\times 10^{3}	$&$	          2.69\times 10^{3}	$&$	          1.91\times 10^{3}	$&$	          2.00\times 10^{3}	$\\
$M_1$         	&$	          1.89\times 10^{3}	$&$	          2.00\times 10^{3}	$&$	          1.78\times 10^{3}	$&$	          1.51\times 10^{3}	$\\
$M_2$         	&$	          5.67\times 10^{3}	$&$	          6.00\times 10^{3}	$&$	          5.35\times 10^{3}	$&$	          4.53\times 10^{3}	$\\
$M_3$         	&$	         -3.78\times 10^{3}	$&$	         -4.00\times 10^{3}	$&$	         -3.57\times 10^{3}	$&$	         -3.02\times 10^{3}	$\\
$A_0/m_{16}$         	&$	2.39	$&$	1.37	$&$	0.03	$&$	1.56	$\\
$\tan\beta$      	&$	47.18	$&$	48.05	$&$	47.93	$&$	47.46	$\\
$m_t$            	&$	174.2	$&$	173.1	$&$	174.2	$&$	173.1	$\\
\hline		  		  		  		  	
$\mu$            	&$	3729	$&$	1935	$&$	2913	$&$	2526	$\\

\hline		  		  		  		  	
$m_h$            	&$	125	$&$	126	$&$	124	$&$	123	$\\
$m_H$            	&$	747	$&$	491	$&$	572	$&$	558	$\\
$m_A$            	&$	742	$&$	491	$&$	568	$&$	554	$\\
$m_{H^{\pm}}$    	&$	753	$&$	500	$&$	580	$&$	567	$\\
		  		  		  		  	
\hline		  		  		  		  	
$m_{\tilde{\chi}^0_{1,2}}$	&$	         895,         3739	$&$	         955,         1935	$&$	         848,         2932	$&$	         709,         2540	$\\

$m_{\tilde{\chi}^0_{3,4}}$	&$	        3742,         4822	$&$	        1936,         5043	$&$	        2935,         4562	$&$	        2543,         3849	$\\

$m_{\tilde{\chi}^{\pm}_{1,2}}$	&$	        3789,         4774	$&$	        1934,         5043	$&$	        2978,         4516	$&$	        2579,         3809	 $\\

$m_{\tilde{g}}$  	&$	7694	$&$	7673	$&$	7266	$&$	6239	$\\
		  		  		  		  	
\hline $m_{ \tilde{u}_{L,R}}$	&$	        7667,         6824	$&$	        8112,         7245	$&$	        7219,         6415	$&$	        6295,         5635	 $\\
                 		  		  		  		  	
$m_{\tilde{t}_{1,2}}$	&$	        5331,         6560	$&$	        5604,         6839	$&$	        5239,         6367	 $&$	        4390,         5370	$\\
                 		  		  		  		  	
\hline $m_{ \tilde{d}_{L,R}}$	&$	        7668,         6814	$&$	        8112,         7236	$&$	        7220,         6406	$&$	        6296,         5628	 $\\
                 		  		  		  		  	
$m_{\tilde{b}_{1,2}}$	&$	        5553,         6526	$&$	        5870,         6870	$&$	        5434,         6333	 $&$	        4591,         5341	$\\
                 		  		  		  		  	
\hline		  		  		  		  	
$m_{\tilde{\nu}_{1,2}}$	&$	4148	$&$	4590	$&$	3870	$&$	3487	$\\
                 		  		  		  		  	
$m_{\tilde{\nu}_{3}}$	&$	3898	$&$	4234	$&$	3641	$&$	3243	$\\
                 		  		  		  		  	
\hline		  		  		  		  	
$m_{ \tilde{e}_{L,R}}$	&$	        4153,         2234	$&$	        4590,         2780	$&$	        3875,         2009	 $&$	        3491,         2068	$\\
                		  		  		  		  	
$m_{\tilde{\tau}_{1,2}}$	&$	           1094,   3875   	$&$	        1140,      4235	$&$	  881,      3620 $&$	    1061,    3225     	 $\\
                		  		  		  		  	
\hline		  		  		  		  	
$\Delta(g-2)_{\mu}$  	&$	  3.11\times 10^{-11}	$&$	  3.36\times 10^{-11}	$&$	  3.71\times 10^{-11}	$&$	  4.97\times 10^{-11}	$\\

$\sigma_{SI}({\rm pb})$	&$	  1.59\times 10^{-11}	$&$	  1.29\times 10^{-9}	$&$	  7.08\times 10^{-11}	$&$	  1.00\times 10^{-10}	$\\

$\sigma_{SD}({\rm pb})$	&$	  4.69\times 10^{-10}	$&$	  1.35\times 10^{-9}	$&$	  1.60\times 10^{-9}	$&$	  2.89\times 10^{-9}	$\\

$\Omega_{CDM}h^{2}$	&$	  6.5	$&$	  2.8	$&$	  0.8	$&$	  4.0	$\\
                		  		  		  		  	
\hline		  		  		  		  	
		  		  		  		  	
$R_{t b \tau}$     	&$	1.02	$&$	1.05	$&$	1.03	$&$	1.04	$\\

\hline
\hline
\end{tabular}
\caption{Benchmark points with good Yukawa unification. All the masses are in units of GeV. Point 1, 3 and 4 are generated using Isajet 7.84 whereas point 2 is from SuSpect 2.41. Point 1 and 2 demonstrates how a small value of $R_{t b \tau}$ yields a Higgs mass $\sim 125$ GeV. Point 3 exhibits stau coannihilation and has a small $R_{t b \tau}$ that agrees with $\Omega h^2 < 1$. Point 4 has $m_{16}\simeq m_{10}$ with good YU. }
\label{tab1}
\end{table}

%%%%%%%%%%%%%%%%%%%%%%%%%%%%%%%%%%%%%%%%%%%%%%%%%%%%%%%%%%%%%%%%%%%%%%%%%%%%%%

%%%%%%%%%%%%%%%%%%%%%%%%%%%%%%%%%%%%%
\section{Conclusion \label{conclusions}}

We have demonstrated how $t$-$b$-$\tau $ YU is consistent with a 125 GeV Higgs. Our analysis is an extension of the {analysis} in ref. \cite{Gogoladze:2011aa} in several ways. We have highlighted the effects of threshold corrections on the bottom quark Yukawa coupling in this model, and discussed the implicit relationship between these corrections and the Higgs mass. We showed that for YU better than $\sim$ 5\%, $M_3 > m_{16}$ {at $M_{{\rm GUT}}$}, {with} $M_3 \gtrsim 2 $ {TeV}. This, in turn, leads to a heavy stop quark, $m_{\tilde {t}_R}\gtrsim4$ TeV. The dominant contribution to the Higgs mass arises from the logarithmic dependence of $m_h$ on the stop quark mass. This leads to the prediction $m_{h} \approx$ 125 GeV, consistent with $t$-$b$-$\tau $ YU better than 5\%.

%We found the contribution of the gluino to be the dominant factor effecting the threshold corrections and the Higgs mass in this model. The effect on the threshold corrections is through the gluino-sbottom loop contribution and on the Higgs mass is through logarithmic dependence on the stop mass (which is a function of the gluino mass).

We also compared our results from two different packages, namely Isajet and SuSpect. We found {good agreement between} the two codes with only a few percent difference between the calculations. One important difference is that Isajet {allows} YU better than 2\%, whereas SuSpect has, at best, 5\% YU. Another notable difference is that SuSpect allows for much smaller values of the Higgs mixing parameter $\mu$. This can have implications for natural SUSY since {smaller} $\mu$ {values are preferred in resolving} the little hierarchy problem.

The two codes also agree well {in their predictions of} sparticle masses. We find that insisting on YU better than 5\% implies that the sparticles are heavy enough to evade observation at {the} current LHC energies, but may be observed during the 14 TeV LHC run. Furthermore, we showed that  $t$-$b$-$\tau$ YU predicts a light CP-odd Higgs boson ($A$).  Restricting to $5\%$ and  better YU yields the following bound on the pseudoscalar mass, 400 GeV$\lesssim M_A \lesssim$  1 TeV. Similarly, the bounds on other sparticle masses are $m_{\tilde{g}} \gtrsim 4 $ TeV, $m_{\tilde{\tau}} \gtrsim 500 $ GeV and $m_{\tilde{\chi}^{\pm}} \gtrsim 2 $ TeV for YU 5\% or better.

Finally, we also tested the implications of YU for direct detection of dark matter. We found that {stau-neutralino} coannihilation can lead to {the} correct {dark matter} relic abundance. Moreover, insisting on YU better than  10\% implies a heavy dark matter candidate $(m_{\tilde{\chi}_1^{0}} \gtrsim 400 \rm \ GeV)$. The {neutralino-nucleon} cross sections are found to be well below the current sensitivity of direct detection experiments.

%%%%%%%%%%%%%%%%%%%%%%%%%%%%%%%%%%%%%%%%%%%%%%
\section*{Acknowledgments}
We would like to thank K.S. Babu, Alexander Pukhov, Shabbar Raza and Zurab Tavartkiladze    for useful discussions.
This work is supported in part by the DOE Grant No. DE-FG02-91ER40626. This work used the Extreme Science
and Engineering Discovery Environment (XSEDE), which is supported by the National Science
Foundation grant number OCI-1053575.

%%%%%%%%%%%%%%%%%%%%%%%%%%%%%%%%%

\end{document}